# Broadband dispersionless polarization rotation with composite chiral metamaterials

Kun Song[*], Changlin Ding, Yahong Liu, Chunrong Luo, and Xiaopeng Zhao

Smart Materials Laboratory, Department of Applied Physics, Northwestern Polytechnical University, Xi'an 710129, P. R. China

E-mail: songkun@nwpu.edu.cn

**Abstract**：We propose a planar composite chiral metamaterial (CCMM) by symmetrically inserting a metallic mesh between two layers of conjugated gammadion resonators. As the elaborate CCMM operates in the frequency region of off-resonance, it therefore presents loss-less and dispersion-free features. It can achieve flat polarization rotation simultaneously accompanied with high transmission and extremely low ellipticity in a broad bandwidth. In the meanwhile, this intriguing CCMM shows more superiorities in polarization rotation power and operating bandwidth than the pure chiral metamaterial just composed of conjugated gammadions. Due to the fascinating properties, the proposed CCMM is greatly appealing for controlling the polarization state of the electromagnetic waves.

## 1. Introduction

An object which cannot be superimposed with its mirror image can be regarded as chiral [1]. In nature, from molecular to crystal, and to polymer, chirality exists ubiquitously, and even the building blocks of life—DNA is also chiral. Substances with chirality can rotate the polarization plane of the electromagnetic waves, which may hold great promises for potential applications in realms such as life science, optoelectronic, and telecommunications, etc [2]. Nevertheless, the normally weak polarization rotation power of naturally chiral materials extremely limits their practical applications, especially in low frequency region and micro-nano devices.

Metamaterials, a kind of manmade materials with tailored electromagnetic properties, provide an efficient route to manipulate electromagnetic wave propagation. With well-designed geometrical structures, metamaterials can accomplish plenty of extraordinary phenomena that do not exist in naturally available materials [3-10]. As an important subset of metamaterials, chiral metamaterials (CMMs) exhibit excellent performances including strong polarization rotation power [11-16], negative refractive index [4,17-20], circular dichroism [18,21,22], asymmetrical transmission [23-28], and so on. In the part few years, utilizing strong resonant responses, the CMMs have realized giant polarization rotation power which is several orders of magnitude larger than that of natural materials, revealing that CMMs are very suitable for polarization manipulation [4,17-20,29-31]. However, it is worthy of note that these resonance-based CMMs inevitably suffer from highly dispersive polarization rotation and high losses at the resonant frequencies [13,18-20]. At the same time, in the off-resonance frequency region, the polarization rotation properties of these CMMs are still influenced by the neighbouring resonances owing to the finite spectral interval between resonances [13,14,18-21]. Therefore, most of the previously reported CMMs cannot realize broadband and constant polarization rotation with high transmission. Recently, much effort has been devoted toward designing and investigating dispersion-free CMMs. Hannam *et al*. [32] first demonstrated a complementary CMM that could achieve dispersionless polarization rotation at the transmission resonance. However, this CMM can only operate in a narrow bandwidth. Then, Park and Li *et al*. [33,34] each proposed different kinds of three-dimensional helical metamaterials which presented nondispersive polarization rotation. Despite the described advantages, these three-dimensional metamaterials are more difficult to fabricate than the planar metamaterials.

In this paper, we propose a novel planar composite CMM (CCMM) composed of conjugated gammadion resonators and subwavelength meshes which presents much stronger polarization rotation power than the pure CMM (PCMM) just consisting of conjugated gammadion resonators. Since the present CCMM operates in the off-resonance frequency region, the dispersion effect due to resonance can thus be effectively suppressed, and the CCMM exhibits low-loss and low-dispersion features. Therefore, it can achieve flat polarization rotation and extremely low circular dichroism in a broadband frequency range. Meanwhile, in the operating region the transmission of the proposed CCMM is over 86%. In addition, compared with the three-dimensional CMMs [22,33,34], the fabrication of the present CCMM is much easier and can be readily extended to optical frequency range, thereby enabling more extensive applications.

## 2. Simulations and experiments

Figure 1 presents the schematic diagram and physical photographs of the proposed CCMM. The considered CCMM consists of a subwavelength mesh symmetrically sandwiched by two layers of conjugated gammadion resonators. Along the $z$ axis, the unit cell is fourfold ($C_4$) symmetry. In our design, both of the dielectric substrates are RT5880, of which the relative permittivity is 2.65 with a dielectric loss tangent of 0.001. The metallic patterns are made of copper with a thickness of 0.035 mm and a conductivity of $5.8 \times 10^7$ S/m. The detailed geometrical parameters of the unit cell are shown in the captions of Fig. 1.

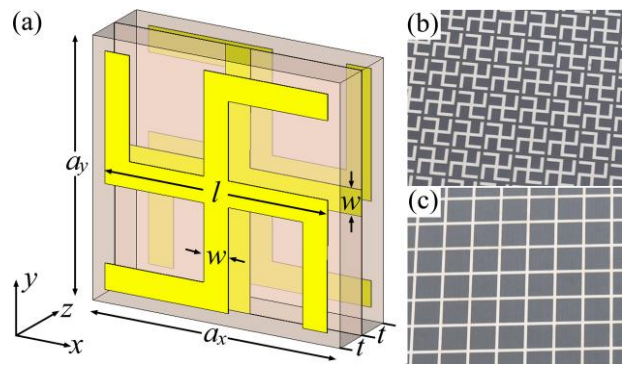

Figure 1 (a) The schematic view of the designed CCMM. The structural dimensions are as follows: $a_x = a_y = 10$ mm, $l = 9$ mm, $w = 1$ mm, and $t = 1.5$ mm. (b) and (c) Photographs of the front and middle layers of the experimental sample.

In order to study the electromagnetic properties of the present CCMM, we carried out numerical simulations and experiments. The simulations were achieved by using the commercial software CST Microwave Studio. In simulations, the unit cell boundary was employed in the *x* and *y* directions and open boundary was used in the *z* direction. As the proposed CCMM is C$_4$ symmetry, the cross transmissions of circularly polarized waves, *i.e.*, from right-handed circularly polarized (RCP) wave to left-handed circularly polarized (LCP) wave and vice versa, are zero. The transmission coefficients of circularly polarized waves can be obtained by the linear co-polarization and cross-polarization transmission coefficients $T_{xx}$ and $T_{yx}$:

$$T_{\pm} = T_{xx} \mp i * T_{yx},$$

where $T_+$ and $T_-$ are the transmission coefficients of the RCP and LCP waves, respectively. Then, the polarization azimuth rotation angle, which reveals the transmission-phase difference between the RCP and LCP waves, is calculated by:

$$\theta = \frac{\arg(T_+) - \arg(T_-)}{2}.$$

And the ellipticity that characterizes the circular dichroism of the transmitted waves is defined as:

$$\eta = \frac{1}{2}\arctan\frac{|T_+|^2 - |T_-|^2}{|T_+|^2 + |T_-|^2}.$$

In experiments, a CCMM sample consisting of $30 \times 30$ unit cells was fabricated. The linear transmission spectra were obtained via an AV 3629 network analyzer with two broadband horn antennas.

At first, we investigate the transmission behaviors of the PCMM just composed of conjugated gammadion resonators, the structural dimensions of which are the same as those of the CCMM. Figure 2 depicts the results of the PCMM. In spite of the discrepancies due to the imperfect fabrication of the sample and measured errors, the measured curves agree well with the simulated ones. The numerical and measured amplitudes of total transmission of the PCMM, $T = (|T_{xx}|^2 + |T_{yx}|^2)^{1/2}$, are shown in Figs. 2(a) and 2(b), respectively. It is obvious that two resonances occur at 7.9 GHz and 18.5 GHz, respectively. And the transmission amplitude is over 85% in the off-resonance frequency region of 10.4 ~ 15.8 GHz. Especially, the transmission amplitudes are as high as 99% at 11.0 and 15.0 GHz. Figures 2(c) and 2(d) portray the simulated and experimental results of the polarization azimuth rotation angle $\theta$, separately. It is found that the curves of the polarization azimuth rotation angle are relatively flat within the band from 11.5 GHz to 13.7 GHz, and the maximum polarization azimuth rotation angle is about 13.7 ° with a variation less than 1 °. The results of the ellipticity $\eta$ are shown in Figs. 2(e) and 2(f). It is known that the CMM will generate a pure polarization rotation effect at $\eta = 0$, *i.e.*, as a linearly polarized incident wave passes through the CMM, the transmitted wave is still linearly polarized but with the polarization plane rotated by an angle of $\theta$ [19]. It can be seen in Figs. 2(e) and 2(f) that the ellipticity $\eta$ is approximatively zero in the frequency range of 11.0 ~ 15.3 GHz, revealing that the proposed PCMM can realize pure polarization rotation in this region. These fascinating properties mentioned above indicate that the PCMM can achieve nearly changeless pure polarization rotation with high transmission in the range of 11.5 ~ 13.7 GHz (labeled by light gray). This arises due to that, since the two resonances are well separated, the losses and dispersive responses resulting from resonances will have much less effect on the electromagnetic properties of the PCMM, thus the PCMM can exhibit high transmission and low-dispersive polarization rotation in the off-resonance region.

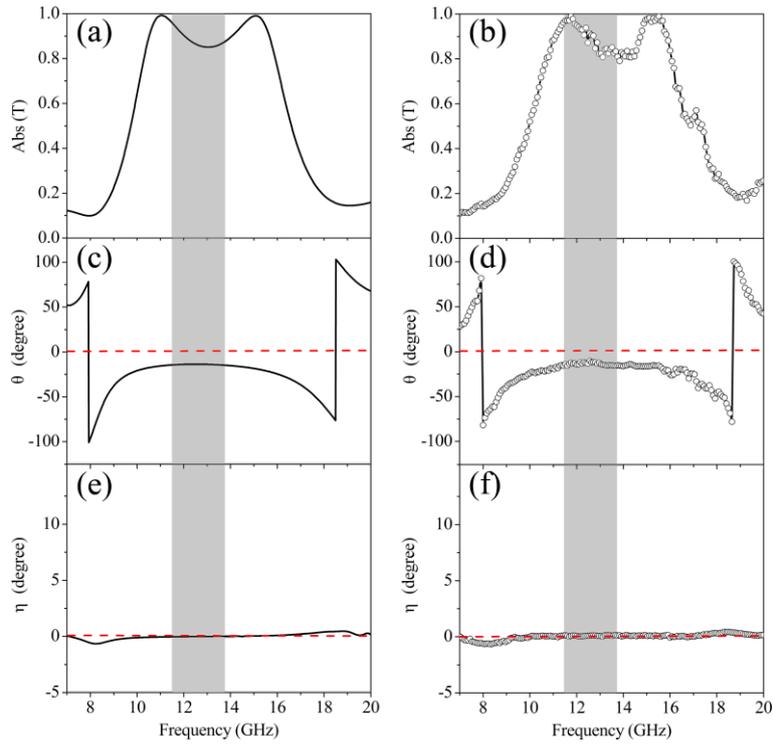

Figure 2. The simulation and measurement results of the PCMM just consisting of conjugated gammadion resonators. (a) and (b) Transmission spectra, (c) and (d) Polarization azimuth rotation angle $\theta$, (e) and (f) Ellipticity $\eta$.

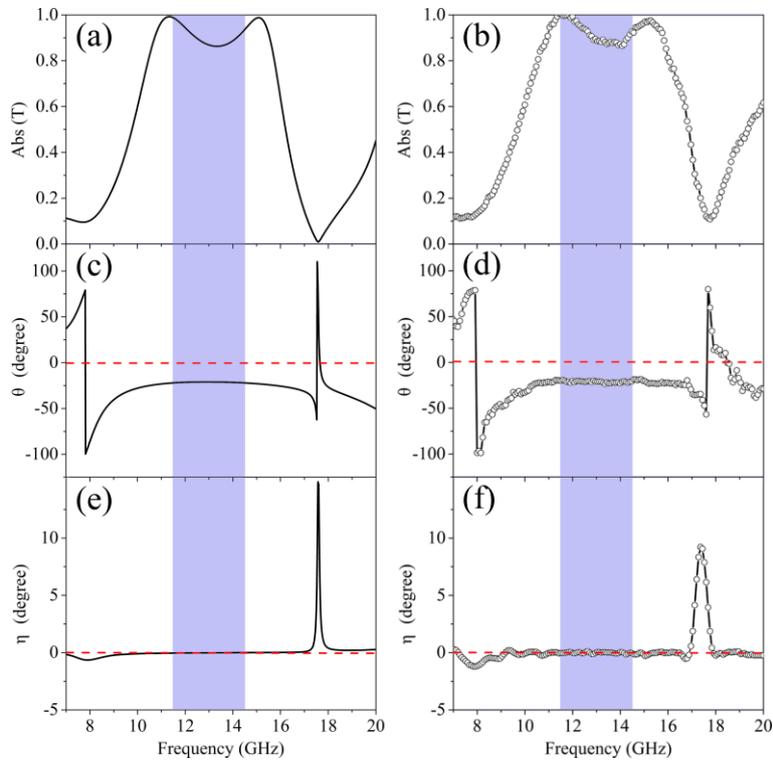

Figure 3. The numerical and measured results of the CCMM. (a) and (b) Transmission spectra, (c) and (d) Polarization azimuth rotation angle $\theta$, (e) and (f) Ellipticity $\eta$.

Although the properties of the PCMM sound interesting, it should be noted that the polarization rotation power of the PCMM is weak. In the previous papers, it has been demonstrated that the CCMM may possess some more excellent electromagnetic properties than the PCMM, such as high figure of merit, enhanced polarization rotation, and angle-insensitivity [35,36]. On the basis of these features, an intriguing CCMM is constructed by symmetrically imbedding the subwavelength meshes between the two layers of conjugated gammadion resonators, as shown in Fig. 1. In Fig. 3, we illustrate the simulation and experimental results of the considered CCMM. It can be seen in Figs. 3(a) and 3(b) that there are also two resonances occurring at 7.8 GHz and 17.6 GHz on the transmission curves of the CCMM, respectively. In comparison with Figs. 2(a) and 2(b), the resonant frequencies of the CCMM show significant red shifts, and the two resonances become close to each other. Moreover, the transmission amplitude of the CCMM is above 86% between 10.6 GHz and 15.6 GHz, a little higher than that of the PCMM. Figures 3(c) and 3(d) plot the results of the polarization azimuth rotation angle $\theta$. It is found that the maximum polarization azimuth rotation angle of the CCMM rises up to 21°, exhibiting much stronger polarization rotation power than that of the PCMM, and the variation of polarization azimuth rotation angle of the CCMM is less than 1° between 11.0 GHz and 14.5 GHz. Meanwhile, the curves of the polarization azimuth rotation angle of the CCMM in the off-resonance region become more flat than those of the aforementioned PCMM. In Figs. 3(e) and 3(f), the values of the ellipticity $\eta$ of the CCMM are kept to be zero within the band from 11.0 GHz to 16.8 GHz, indicating that a pure polarization rotation effect is accomplished in this region. Furthermore, at the second resonance, the ellipticity $\eta$ radically increases. According to the intriguing manifestations mentioned above, in the off-resonance frequency range of 11.5 ~ 14.5 GHz (labeled by light purple) where we are most interested, the CCMM can achieve high-performance and broadband dispersionless polarization rotation. Compared the CCMM with the PCMM, we can find that the CCMM shows more advantages in the operating bandwidth and polarization rotation power than the PCMM.

Figures 4(a) and 4(b) show the differentials of transmission amplitude and polarization azimuth rotation angle to frequency for the CCMM and PCMM, respectively. It shows that in Fig. 4(a) the maximum of the differential of transmission amplitude to frequency for the CCMM is less than that of the PCMM. In Fig. 4(b), it is significant that the change of the differential of polarization azimuth rotation angle to frequency for the CCMM is much smaller than that of the PCMM. These facts further confirm that the elaborately designed CCMM exhibits more excellent dispersionless features than the PCMM and can achieve flat pure polarization rotation in a more broadband range. Therefore, the CCMM may be a good candidate to realize highly efficient and dispersion-free polarization rotation in a broad bandwidth.

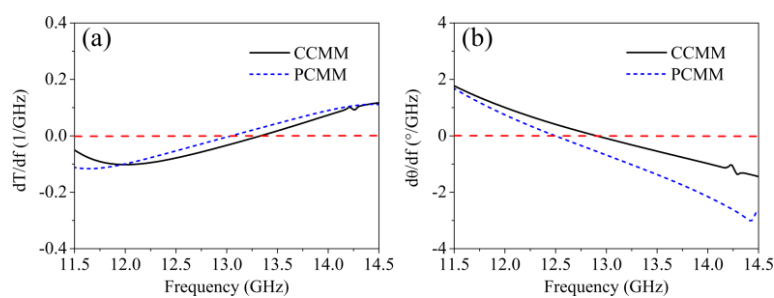

Figure 4. Differentials of (a) transmission amplitude and (b) polarization azimuth rotation angle to frequency for the CCMM and PCMM.

## 3. Discussions

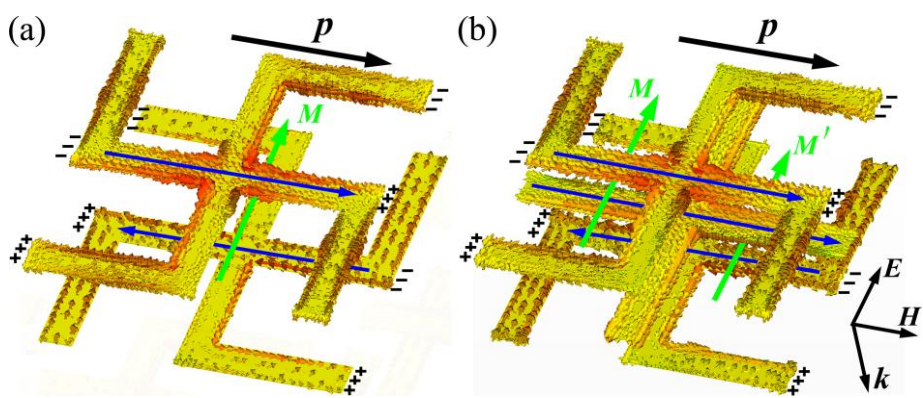

Figure 5. Surface current distributions of (a) PCMM and (b) CCMM at 13.0 GHz and 13.3 GHz, respectively.

To elucidate the underlying physics of the enhanced polarization rotation, we examine the surface current distributions of the PCMM and CCMM in the off-resonance frequency region, as shown in Fig. 5. Figure 5(a) shows the surface current distributions of the PCMM at 13.0 GHz. It is seen that the external *E*-field induces circular currents on the U-shaped parts of the gammadions, and leads to the antiparallel currents (blue arrows) on the center parts of the front and back gammadions, which induces a magnetic moment *M* parallel to the *E*-field direction. Moreover, the external *E*-field and *H*-field will also induce positive and negative charges accumulating on the opposite sides of the gammadions. The charges vary with time, finally generating an electric dipole *P* parallel to the *H*-field direction. This is the origin of the chirality of the PCMM. In Fig. 5(b), a similar physical mechanism occurs for the proposed CCMM as well. However, except for the magnetic moment *M*, the antiparallel currents on the center parts of the embedded mesh and back layer gammadion induce another magnetic moment *M'*, which is in the same direction as the magnetic moment *M* and results in the enhancement of the chirality of the CCMM. Additionally, it has also been confirmed that chirality will be effectively enhanced as the two resonances become close to each other [20,36]. In comparing Fig. 3 with Fig. 2, since the two resonances of the CCMM become closer, the polarization rotation power of the CCMM is therefore increased further.

## 4. Conclusions

In conclusion, we have numerically and experimentally demonstrated a planar CCMM obtained by the combination of the conjugated gammadion resonators and a subwavelength mesh. Due to the proposed CCMM operating in the off-resonance region, the losses and dispersive responses resulting from resonances are effectively suppressed. Thus, the CCMM presents low-loss and low-dispersive characteristics. The results show that the intriguing CCMM can achieve broadband and nearly constant pure polarization rotation with high efficiency. Meanwhile, the polarization rotation power and operating bandwidth of the present CCMM are significantly enhanced comparing with the PCMM. And the mechanism of the enhanced polarization rotation power of the CCMM is physically explained. Additionally, the fabrication of this planar CCMM is much easier than that of the three-dimensional

CMMs. The idea of constructing CCMM provides us a flexible route to design despersionless CMMs. With the unique performances, the present CCMM may greatly benefit the potential applications in microwave, terahertz, and optical devices.

**Acknowledgements**

This work was supported by the National Natural Science Foundation of China (Grant Nos. 11174234, 51272215, 11204241, 11404261), the National Key Scientific Program of China (under project No. 2012CB921503) and the Northwestern Polytechnical University Scientific Research Allowance (No. G2015KY0302).